\documentstyle[12pt,epsf]{article}
\newcommand{\lsim}{\mathrel{\raisebox{-.6ex}{$\stackrel{\textstyle<}{\sim}$}}}
\newcommand{\gsim}{\mathrel{\raisebox{-.6ex}{$\stackrel{\textstyle>}{\sim}$}}}
\def\eps{\epsilon}

\def\beq{\begin{equation}}
\def\eeq{\end{equation}}
\def\bea{\begin{eqnarray}}
\def\eea{\end{eqnarray}}

\def\half{\frac{1}{2}}

\begin{document}

\thispagestyle{empty}

\font\fortssbx=cmssbx10 scaled \magstep2
\hbox to \hsize{
\hbox{\fortssbx University of Wisconsin - Madison}
      \hfill$\vcenter{
\hbox{\bf MADPH-98-1063}
\hbox{\bf HAWAII-511-905-98}
\hbox{\bf VAND-TH-98-08}
\hbox{\bf AMES-HET-98-09}
       \hbox{June 1998}}$ }

\vspace{.5in}

\begin{center}
{\bf BI-MAXIMAL MIXING OF THREE NEUTRINOS}
\\
\vskip 0.7cm
{V. Barger$^1$, S. Pakvasa$^2$, T.J. Weiler$^3$, and K. Whisnant$^4$}
\\[.1cm]
$^1${\it Department of Physics, University of Wisconsin, Madison, WI
53706, USA}\\
$^2${\it Department of Physics and Astronomy, University of Hawaii,
Manoa, HI 96822, USA}\\
$^3${\it Department of Physics and Astronomy, Vanderbilt University,
Nashville, TN 37235, USA}\\
$^4${\it Department of Physics and Astronomy, Iowa State University,
Ames, IA 50011, USA}\\
\end{center}

\smallskip

\begin{abstract}

We show that if the solar and atmospheric data are both described by
maximal vacuum oscillations at the relevant mass scales then there
exists a unique mixing matrix for three neutrino flavors. The solution
necessarily conserves $CP$ and automatically implies that there is no
disappearance of atmospheric $\nu_e$, consistent with indications from
the Super-Kamiokande experiment. We also investigate the consequences
for three-neutrino mixing if the solar and atmospheric oscillations
exhibit mixing that is large but not maximal. For non-maximal mixing
$\nu_e\leftrightarrow\nu_\tau$ and $\nu_e\leftrightarrow\nu_\mu$
oscillations are predicted that may be observable in future
long-baseline experiments.

\end{abstract}

\thispagestyle{empty}
\newpage

{\it Introduction.} It is now understood that neutrino oscillations can
describe the data on the solar neutrino deficit
\cite{SSM,solar,newsolar}, the atmospheric neutrino anomaly
\cite{atmos,SuperK,oldatmos}, and the results from the LSND experiment
\cite{LSND}, only if a sterile neutrino is introduced
\cite{models,bgg,bww97,gmnr98,bpww}. However, because confirmation of
the LSND results awaits future experiments \cite{LSNDconfirm}, a
conservative approach is to assume that oscillations need only account
for the solar and atmospheric data; also, it is possible to introduce
new lepton-flavor changing operators with coefficients small enough to
evade present exclusion limits, but large enough to explain the small
LSND amplitude \cite{jm98}. Since a comparison of the theoretical
expectations \cite{flux} with the Super-Kamiokande (SuperK) atmospheric
data suggests maximal $\nu_\mu$ oscillations with mass-squared difference
$\delta m^2_{atm} \approx 5\times10^{-3}$~eV$^2$ \cite{SuperK}, and in
light of the recent data from Super-Kamiokande \cite{newsolar} favoring
vacuum long-wavelength oscillations \cite{bpw81,justso,vlw}
with $\delta m^2_{sun} \approx 10^{-10}$~eV$^2$ and large mixing as a
description of the solar neutrino deficit, it is
instructive to determine how maximal or near-maximal solar and
atmospheric vacuum oscillations may be accommodated within a
three-neutrino universe.

In this letter we use unitarity constraints to derive conditions on the
three-neutrino mixing matrix under the assumption that both solar and
atmospheric neutrinos undergo maximal mixing {\it in vacuum}.  By
maximal mixing we mean that the disappearance probabilities are
equivalent to those for maximal two-neutrino mixing at the relevant mass
scales. We find that there exists a unique three-neutrino mixing
solution, up to trivial sign ambiguities, which conserves $CP$ and has
no oscillations of atmospheric $\nu_e$, consistent with indications from
the Super-Kamiokande experiment \cite{SuperK}. This mixing matrix
predicts that solar $\nu_e$ oscillate maximally into equal numbers of
$\nu_\mu$ and $\nu_\tau$. We also investigate the consequences for
three-neutrino mixing if the solar and atmospheric oscillations are not
maximal. We find that for near-maximal solar and atmospheric neutrino
mixing the $\nu_\mu \rightarrow \nu_e$ and $\nu_e \rightarrow \nu_\tau$
oscillation amplitudes for atmospheric and long-baseline experiments
are approximately equal, and are limited by the deviation of the solar
$\nu_e$ disappearance amplitude from unity. Also, for any non-maximal
solar and atmospheric neutrino mixing there must exist $\nu_e
\rightarrow \nu_\tau$ oscillations that may be visible in future
long-baseline experiments.

{\it Oscillation probabilities.} We begin our analysis with the
survival probability for neutrino oscillations in a vacuum
\cite{VBreal}
\begin{equation}
P(\nu_\alpha\to \nu_\alpha) = 1
- 4 \sum_{k<j} P_{\alpha j} P_{\alpha k} \sin^2 \Delta_{jk} \,,
\label{oscprob}
\end{equation}
where $P_{\alpha j}\equiv |U_{\alpha j}|^2$ is the probability to find
the $\alpha$-flavor state in the $j$th mass state (or vice versa),
$U$ is the neutrino mixing matrix (in the basis where the charged-lepton
mass matrix is diagonal),
$\Delta_{jk} \equiv \delta m_{jk}^2 \,L/4E =
1.27 (\delta m^2_{jk}/{\rm eV}^2) (L/{\rm km})/(E/{\rm GeV})$,
$\delta m^2_{jk}\equiv m^2_j-m^2_k$,
and the sum is over all $j$ and $k$, subject to $k<j$. The
matrix elements $U_{\alpha j}$ are the mixings between
the flavor ($\alpha=e,\mu,\tau$) and the mass ($j=1,2,3$)
eigenstates. We assume a neutrino mass spectrum with mass eigenvalues
$m_1, m_2 \ll m_3$, where the solar oscillations are driven by
$\Delta_{21} \equiv \Delta_{sun}$ and the atmospheric oscillations are
driven by $\Delta_{31} \simeq \Delta_{32} \equiv \Delta_{atm}$. However,
only the form of the neutrino mass matrix depends on the mass hierarchy
assumption. All of our other conclusions require only the more general
assumption that $|\Delta_{sun}| = |\Delta_{21}| \ll |\Delta_{31}| \simeq
|\Delta_{32}| = |\Delta_{atm}|$, which also could apply, e.g., to the
case $m_3 < m_1 \simeq m_2$, i.e., the solar neutrino vacuum
oscillations are driven by a small mass splitting between the two larger
masses.

The off-diagonal oscillation probabilites of this model are
\begin{eqnarray}
P(\nu_e\rightarrow\nu_\mu) &=&
4\,P_{e3} P_{\mu 3} \sin^2\Delta_{atm}
-4 Re\{ U_{e1} U_{e2}^* U_{\mu 1}^* U_{\mu 2}\} \sin^2\Delta_{sun}
-2\,J\sin 2\Delta_{sun}\,,
\label{pemu} \\
P(\nu_e\rightarrow\nu_\tau) &=&
4\,P_{e3} P_{\tau 3} \sin^2\Delta_{atm}
-4 Re\{U_{e1} U_{e2}^* U_{\tau 1}^* U_{\tau 2}\} \sin^2\Delta_{sun}
+2\,J\sin 2\Delta_{sun}\,,
\label{petau} \\
P(\nu_\mu\rightarrow\nu_\tau) &=&
4\,P_{\mu 3} P_{\tau 3}\sin^2\Delta_{atm}
-4 Re\{ U_{\mu 1} U_{\mu 2}^* U_{\tau 1}^* U_{\tau 2}\} \sin^2\Delta_{sun}
-2\,J\sin 2\Delta_{sun}\,,
\label{pmutau}
\end{eqnarray}
where the $CP$-violating ``Jarlskog invariant'' \cite{jarlskog}
is $J = Im\{ U_{e2} U_{e3}^* U_{\mu 2}^* U_{\mu 3}\}$.
The $CP$-odd term changes sign under reversal of the oscillating flavors.
The solar amplitudes, quartic in the $U_{\alpha j}$'s, may themselves
be expressed in terms of the $P_{\alpha j}$ \cite{kpbook,ww98}, as may
the $CP$-odd amplitude $J$ \cite{ww98,HnJ}.

We note that the $CP$-violating probability at the atmospheric scale is
suppressed to order $\delta m^2_{sun}/\delta m^2_{atm}$, the leading
term having canceled in the sum over the two light-mass states.  Thus,
$P(\nu_\alpha \rightarrow \nu_\beta) = P(\nu_\beta \rightarrow
\nu_\alpha)$ at the atmospheric scale. With MSW-enhanced solutions
to the solar deficit, $\sin 2\Delta_{sun}$ is effectively replaced by
its average value of zero since $\Delta_{sun} \gg 1$ in this case. Only
for the vacuum long-wavelength solution to the solar neutrino anomaly
is there hope to measure the $CP$-odd term.


We will use ``amplitude'' to denote the coefficients of the oscillating
factors. The amplitudes of interest are:
\bea
 A^{e\not e}_{sun} &=& 4 \, P_{e1} P_{e2} \leq (1 - P_{e 3})^2 \,,
\label{sunamp}\\
A^{\mu\not\mu}_{atm} &=& 4 \, P_{\mu 3} (P_{\mu 1} + P_{\mu 2})
 = 4 \, P_{\mu 3} (1 - P_{\mu 3}) \,,
\label{atmamp}\\
A^{e\not e}_{atm} &=& 4 \, P_{e 3} (1 - P_{e 3}) \,,
\label{atmee}\\
A^{\mu e}_{atm} &=& 4 \, P_{e 3} P_{\mu 3} \leq (1 - P_{\tau 3})^2 \,,
\label{atmue}\\
A^{\mu\tau}_{atm} &=& 4 \, P_{\mu 3} P_{\tau 3} \leq (1 - P_{e 3})^2 \,,
\label{atmutau}\\
A^{e\tau}_{atm} &=& 4 \, P_{e 3} P_{\tau 3} \leq (1 - P_{\mu 3})^2 \,.
\label{atmetau}
\eea
In a three-neutrino model
\bea
A^{\mu\not\mu}_{atm} &=& A^{\mu e}_{atm} + A^{\mu\tau}_{atm} \,,
\;\; {\rm and}\\
A^{e\not e}_{atm} &=& A^{\mu e}_{atm} + A^{e\tau}_{atm} \,.
\eea
The equality in Eq.~(\ref{sunamp}) results from the unitarity condition
$U_{\mu 1}^* U_{\mu 2}+U_{\tau 1}^* U_{\tau 2}= - U_{e 1}^* U_{e 2}$.
The inequalities in Eqs.~(\ref{sunamp}), (\ref{atmue})--(\ref{atmetau})
are obtained by maximizing the amplitude under the unitarity constraints
\bea
P_{e 3} + P_{\mu 3} + P_{\tau 3} &=& 1 \,,
\;\; {\rm and}
\label{3unitarity}\\
P_{e 1} + P_{e 2} + P_{e 3} &=& 1\,.
\label{eunitarity}
\eea
Thus, the amplitudes depend on just three independent $P_{\alpha j}$'s,
and the atmospheric amplitudes depend on just two of the
probabilities chosen from $P_{e 3}$, $P_{\mu 3}$, and $P_{\tau 3}$.

{\it Maximal solar and atmospheric oscillations.} We first explore a
solution for which both solar $\nu_e$ and atmospheric $\nu_\mu$ data
can be described by maximal oscillations at the relevant $\delta m^2$
scales, as suggested by the SuperK data \cite{newsolar,SuperK}.
Maximal atmospheric oscillations require $P_{\mu 3} = {1\over 2}$.
If solar oscillations are also required to be maximal, then
$P_{e3}=0$ and $P_{e1} = P_{e2} = {1\over 2}$. Up to trivial
sign changes, unitarity then uniquely specifies the remaining
elements of $U$:
\beq
\left( \begin{array}{c} \nu_e \\ \nu_\mu \\ \nu_\tau \end{array} \right)
= U \left( \begin{array}{c} \nu_1 \\ \nu_2 \\ \nu_3 \end{array} \right)
= \left( \begin{array}{ccc}
 {1\over\sqrt2} & -{1\over\sqrt2} & 0 \\
 {1\over2}      &       {1\over2} & -{1\over\sqrt2} \\
 {1\over2}      &       {1\over2} & {1\over\sqrt2} \\
\end{array} \right)
\left( \begin{array}{c}
\nu_1 \\ \nu_2 \\ \nu_3
\end{array} \right) \;.
\label{U}
\eeq
%
%
With the mixing matrix of Eq.~(\ref{U}),
the off-diagonal oscillation probabilities in
Eqs.~(\ref{pemu})--(\ref{pmutau}) are given by
\begin{eqnarray}
P(\nu_e\rightarrow\nu_\mu) &=&
{1\over2} \sin^2\Delta_{sun} \,,
\label{pemu2} \\
P(\nu_e\rightarrow\nu_\tau) &=&
{1\over2} \sin^2\Delta_{sun} \,,
\label{petau2} \\
P(\nu_\mu\rightarrow\nu_\tau) &=&
\sin^2\Delta_{atm} - {1\over4} \sin^2\Delta_{sun} \,.
\label{pmutau2}
\end{eqnarray}

The zero $U_{e3}$ element implies that $CP$ is conserved, $J$ is zero,
and the elements of $U$ are real. It also implies that the only
oscillations at the $\Delta_{atm}$ scale are
$\nu_\mu\rightarrow\nu_\tau$. The solar $\nu_e$ oscillations are 50\%
into $\nu_\mu$ and 50\% into $\nu_\tau$, although the oscillation
flavors are not distinguishable because they have the same neutral
current interactions and there is not enough energy to produce charge
current events; however, detection of neutrinos from active galactic
nuclei and gamma ray burst sources could distinguish the $\tau$-neutrino
flavor from $\nu_e$ and $\nu_\mu$ \cite{halzen}. The mixing
matrix for this bi-maximal scenario is similar to that obtained in
Refs.~\cite{fritzsch} and \cite{fukugita}, which give maximal solar
oscillations ($A^{e\not e}_{sun}=1$) and large but not maximal mixing of
atmospheric neutrinos ($A^{\mu\not\mu}_{atm}={8\over9}$). All of these
scenarios also have $U_{e3}=0$, which implies maximal solar mixing and
only $\nu_\mu\rightarrow\nu_\tau$ oscillations at the $\Delta_{atm}$
scale.

A Majorana mass matrix or a hermitian Dirac mass matrix can be diagonalized
by a single unitary matrix.  For these two cases, we may therefore invert
the process to obtain $M$ in the flavor basis from the mass eigenvalues $m_j$
and the mixing matrix $U$.
For the Majorana case $M_{Maj}=U M_{diag} U^T$, while
for the hermitian Dirac case $M_{Dirac}=U M_{diag} U^{\dag}$.
The individual matrix elements are
$M_{\alpha\beta}=\sum_j m_j U_{\alpha j} U_{\beta j}$ and
$M_{\alpha\beta}=\sum_j m_j U_{\alpha j} U^*_{\beta j}$,
respectively.  For real $U$, as here, there is no difference
between these two cases.
Reconstruction of the mass matrix from $U$ in Eq.~(\ref{U})
gives
\beq
M = m \left[
\left( \begin{array}{ccc}
0 &  0 & 0 \\
0 &  1 & -1 \\
0 & -1 & 1 \\
\end{array} \right)
+
\left( \begin{array}{ccc}
2\eps   & \delta & \delta \\
\delta  & \eps   & \eps \\
\delta  & \eps   & \eps \\
\end{array} \right)
\right]
\,,
\label{m}
\eeq
in the ($\nu_e$, $\nu_\mu$, $\nu_\tau$) basis,
where the parameters of $M$ are related to the mass eigenvalues as
\begin{eqnarray}
m &=& m_3/2 \,,
\label{em} \\
\eps &=& (m_2 +m_1)/4m \,,
\label{eps} \\
\delta &=&\sqrt{2}\,(m_1-m_2)/4m \,.
\label{delta}
\end{eqnarray}
Individual field redefinitions $\psi_\alpha\rightarrow -\psi_\alpha$ and
$\psi_j\rightarrow -\psi_j$ allow one to change the signs of individual
rows and columns of $M$. We remind the reader that the form of $M$
derived here applies in the basis where the charged-lepton
mass matrix is diagonal.

The two relevant mass-squared differences are
\beq
\delta m^2_{atm} \simeq 4 \, m^2 \,, \qquad
\delta m^2_{sun} = 8 \, \sqrt2 \, \delta \, \eps \, m ^2 \,.
\label{dm2}
\eeq
The atmospheric oscillation scale requires $m \approx
0.035$~eV. The vacuum long-wavelength solar solution then requires
$|\delta\,\eps| \approx 10^{-8}$.  Both $\delta$ and $\eps$ must be
nonzero to generate a nonzero $\delta m^2_{sun}$. We reiterate that only
the form of the mass matrix in Eq.~(\ref{m}) depends on the assumption
$m_1,m_2 \ll m_3$; the oscillation probabilites hold for the more
general relation $|\Delta_{sun}| \ll |\Delta_{atm}|$.

{\it Near-maximal mixing.} We next consider having both solar and
atmospheric neutrino mixing close to maximal ($A^{\mu\not\mu}_{atm}
\approx 1$ and $A^{e\not e}_{sun} \approx 1$). Parametrizing the real mixing
matrix in terms of Euler angles
\beq
\left( \begin{array}{c} \nu_e \\ \nu_\mu \\ \nu_\tau \end{array} \right)
= U \left( \begin{array}{c} \nu_1 \\ \nu_2 \\ \nu_3 \end{array} \right)
= \left( \begin{array}{ccc}
 c_1 c_3                & -c_1 s_3               & s_1 \\
 s_1 c_2 c_3 + s_2 s_3  & -s_1 c_2 s_3 + s_2 c_3 & -c_1 c_2 \\
-s_1 s_2 c_3 + c_2 s_3  &  s_1 s_2 s_3 + c_2 c_3 & c_1 s_2 \\
\end{array} \right)
\left( \begin{array}{c}
\nu_1 \\ \nu_2 \\ \nu_3
\end{array} \right) \,,
\label{U2}
\eeq
where $c_j \equiv \cos\theta_j$ and $s_j \equiv \sin\theta_j$, the
maximal solar and atmospheric mixing of Eq.~(\ref{U}) corresponds to
$\theta_1 = 0$, $\theta_2 = {\pi\over4}$, and $\theta_3 = {\pi\over4}$.
Then for near-maximal mixing we take
\beq
\theta_1 = \eps_1 \,,\qquad
\theta_2 = {\pi\over4} + \eps_2 \,,\qquad
\theta_3 = {\pi\over4} + \eps_3 \,,
\label{theta}
\eeq
where the $\eps_j$ are small. The corresponding oscillation amplitudes are
\begin{eqnarray}
A^{e\not e}_{sun} &\simeq& 1 - 2 \, \eps_1^2 - 4 \, \eps_3^2 \,, \\
A^{\mu\not\mu}_{atm} &\simeq& 1 - 4 \, \eps_2^2 \,, \\
A^{\mu e}_{atm} &\simeq& A^{e\tau}_{atm} \simeq
{1\over2} A^{e\not e}_{atm} \simeq 2 \, \eps_1^2 \,.
\end{eqnarray}
Here atmospheric $\nu_\mu$ disappearance is determined solely by
$\eps_2$. The solar $\nu_e$ disappearance and atmospheric and
long-baseline $\nu_\mu\rightarrow\nu_e$ and $\nu_e\rightarrow\nu_\tau$
oscillation amplitudes must satisfy the conditions
\beq
\half A^{e\not e}_{atm} \simeq A^{\mu e}_{atm}
\simeq A^{e\tau}_{atm} \lsim 1 - A^{e\not e}_{sun} \,.
\label{equal}
\eeq
Hence for near-maximal mixing, $\nu_\mu\rightarrow\nu_e$ and
$\nu_e\rightarrow\nu_\tau$ oscillations in atmospheric and
long-baseline experiments are small and approximately equal, and they
are constrained by how close the solar mixing is to maximal. The
atmospheric $\nu_\mu$ disappearance amplitude is independent of the
other oscillation amplitudes in this limit.

{\it Non-maximal mixing.} We also investigate the more general case of
large but non-maximal solar neutrino mixing, which corresponds to
nonzero $U_{e3}$, and large but non-maximal atmospheric neutrino
mixing. A geometric expression for $J$ \cite{jarlskog,ww98} equates its
magnitude to the area of a triangle with sides which may be taken to be
$|U_{e3} U_{\mu 3}|$, $|U_{e2} U_{\mu 2}|$, and $|U_{e1} U_{\mu 1}|$.
Since $|U_{e3}|$ is known to be small, we use Eq.(\ref{U}) as an
approximate guide for $U$ and obtain $|J| \lsim |U_{e3}|/8$ as the order
of magnitude for the area of the triangle.  This bound on $|J|$ is
sufficiently small that we will set $|J|$ to zero and continue to work
with real $U$ and three independent parameters in the mixing matrix.  If
the oscillation amplitudes $A^{e\not e}_{sun}$ and
$A^{\mu\not\mu}_{atm}$ are determined, e.g., from experiment, then there
remains only one free parameter, e.g., $A^{e\not e}_{atm}$. Although
pure $\nu_\mu\leftrightarrow\nu_e$ oscillations of atmospheric neutrinos
are strongly disfavored, modest amounts of $\nu_e$ disappearance have
not yet been ruled out. Therefore it is worthwhile to consider the
effects of a nonzero $U_{e3}$ when the approximations of the previous
section are no longer valid.

Given the inputs $A^{e\not e}_{sun}$, $A^{\mu\not\mu}_{atm}$, and
$A^{e\not e}_{atm}$, the three independent parameters of the matrix $U$
can be determined up to some two-fold ambiguities, only one of which
has phenomenological consequences for vacuum oscillations. For example,
knowing the atmospheric $\nu_\mu$ disappearance amplitude in
Eq.~(\ref{atmamp}) determines $U_{\mu 3}=\pm\sqrt{P_{\mu 3}}$:
\beq
P_{\mu 3} =
{1\over2} \left[
1 \mp \sqrt{1 - A^{\mu\not\mu}_{atm}} \,
\right] \,,
\label{Umu3}
\eeq
where the $-$ ($+$) means that $\nu_\mu$ is more closely associated with
$\nu_2$ ($\nu_3$). Knowing the atmospheric $\nu_e$ disappearance
amplitude in Eq.~(\ref{atmee}) determines $U_{e 3}=\pm\sqrt{P_{e 3}}$:
\beq
P_{e 3} =
{1\over2} \left[
1 - \sqrt{1 - A^{e\not e}_{atm}} \,
\right] \,.
\label{Ue3}
\eeq
In Eq.~(\ref{Ue3}) we have chosen the solution which gives
$P_{e3} \leq {1\over2}$ since $P_{e3} > {1\over2}$ implies that
the solar $\nu_e$ disappearance amplitude $A^{e\not e}_{sun}$ must be
less than ${1\over4}$, which for the long-wavelength solar solution is
disfavored by the data. Then the unitarity relation in
Eq.~(\ref{3unitarity}) yields
\beq
P_{\tau 3} = {1\over2} \left[ 
\sqrt{1-A^{e\not e}_{atm}} \pm \sqrt{1-A^{\mu\not\mu}_{atm}} \,
\right] \,,
\label{Utau3}
\eeq
where the $\pm$ in Eq.~(\ref{Utau3}) is correlated with the $\mp$ in
Eq.~(\ref{Umu3}). Also, Eq.~(\ref{sunamp}) and the unitarity relation
in Eq.~(\ref{eunitarity}) imply
\begin{eqnarray}
P_{e1} &=& {1\over2} \left[ 1 - P_{e3}
+ \sqrt{(1 - P_{e3})^2 - A^{e\not e}_{sun}} \, \right] \,,
\label{Ue1} \\
P_{e2} &=& {1\over2} \left[ 1 - P_{e3}
- \sqrt{(1 - P_{e3})^2 - A^{e\not e}_{sun}} \, \right] \,,
\label{Ue2}
\end{eqnarray}
where we have assumed that $\nu_e$ is more closely associated with
$\nu_1$ than with $\nu_2$ (the other possibility, that $\nu_e$ is more
closely associated with $\nu_2$, has similar vacuum oscillation
phenomenology). In principle, there is a constant term in the solar
oscillation probability $2\,P_{e3}(1-P_{e3}) = \half A^{e\not e}_{atm}$,
from averaging of the $\sin^2\Delta_{atm}$ term.  However, if $A^{e\not
e}_{atm}$ is small, as suggested by the non-observation of $\nu_e$
oscillations in the atmospheric data, this term can be ignored to first
approximation \cite{osland}. The remaining elements of $U$ may also be
determined from unitarity (e.g., by using the values of $P_{\mu 3}$,
$P_{e3}$, $P_{\tau 3}$, $P_{e1}$, and $P_{e2}$ and the form of
Eq.~(\ref{U2}) to determine the three Euler mixing angles).

Once the oscillation parameters are fixed (up to the two-fold ambiguity
in Eq.~(\ref{Umu3})), other oscillation amplitudes may be calculated;
for example, Eqs.~(\ref{Umu3}) and (\ref{Ue3}) combined with
Eq.~(\ref{atmue}) yield
\beq
A^{\mu e}_{atm} = \left[1-\sqrt{1 - A^{e\not e}_{atm}} \, \right]
\left[1 \mp \sqrt{1 - A^{\mu\not\mu}_{atm}} \, \right] \,,
\label{amue}
\eeq
where the $\mp$ is correlated with the $\mp$ in Eq.~(\ref{Umu3}).
Contours of $A^{\mu e}_{atm}$ in the
($A^{\mu\not\mu}_{atm}$,$A^{e\not e}_{atm}$) plane are shown in
Figs.~1(a) and 1(b). Once $A^{\mu\not\mu}_{atm}$ and $A^{e\not e}_{atm}$
are measured, the value of $A^{\mu e}_{atm}$ is predicted.

In principle the three inputs can be determined by the solar and
atmospheric neutrino data. For instance, $A^{e\not e}_{sun}$ is directly
determined from vacuum long-wavelength fits to the solar neutrino
spectrum. Also, the two independent atmospheric oscillation amplitudes
can be found from measurements of
%
\beq
N_\mu/N^o_\mu = (1 - <S> \, A^{\mu\not\mu}_{atm})
+ r \, <S> \, A^{\mu e}_{atm} \,,
\eeq
and
\beq
N_e/N^o_e = (1 - <S> \, A^{e\not e}_{atm})
+ r^{-1} \, <S> \, A^{\mu e}_{atm}\,,
\eeq
where $N_e^o$ and $N_\mu^o$ are the expected number of atmospheric $e$
and $\mu$ events, $r\equiv N^o_e/N^o_\mu \sim \half$, and $<S>$ is
$\sin^2\Delta_{atm}$ averaged over the appropriate atmospheric neutrino
spectrum.

The existence of a real solution to Eqs.~(\ref{Ue1}) and (\ref{Ue2})
requires
\beq
P_{e3} \leq 1 - \sqrt{A^{e\not e}_{sun}} \,.
\label{pe3lim}
\eeq
The limit in Eq.~(\ref{pe3lim}) becomes stringent when
$A^{e\not e}_{sun}$ is close to unity. Current data suggest
$A^{e\not e}_{sun} \geq 0.6$, which gives $P_{e3} \leq 0.23$ and hence
$A^{e\not e}_{atm} \leq 0.7$. These
limits will improve as more solar and atmospheric data become available.
As a check, note that in Eq.~(\ref{pe3lim}) maximal mixing of solar
neutrinos ($A^{e\not e}_{sun}=1$) automatically leads to
$|U_{e3}| = A^{\mu e}_{atm} = A^{e\tau}_{atm} = 0$, i.e., all
atmospheric oscillations are $\nu_\mu\rightarrow\nu_\tau$.
{}From our approximate bound for $|J|$ when $U_{e3}$ is small, we
obtain $|J| \lsim 0.06$, which verifies that $CP$-violation effects will
be small.

Another limit on $P_{e3}$ comes from the CHOOZ reactor experiment
\cite{CHOOZ} that measures $\bar\nu_e$ disappearance
\beq
A^{e\not e}_{atm} \lsim 0.2 \,,
\label{choozlimit}
\eeq
which applies for $\delta m^2_{atm} \gsim 2\times10^{-3}$~eV$^2$. When
combined with Eq.~(\ref{Ue3}), Eq.~(\ref{choozlimit}) gives the bound
\beq
P_{e3} \lsim 0.05 \,.
\label{pe3lim2}
\eeq
However, for $\delta m^2_{atm} < 10^{-3}$~eV$^2$ there is no
limit at all. The proposed experiment to detect reactor anti-neutrinos
in the BOREXINO detector \cite{borexino} can test a value of
$A^{e\not e}_{atm}$ as low as 0.4 (corresponding to a limit $P_{e3}
\leq 0.18$) for $\delta m^2_{atm} \gsim 10^{-6}$~eV$^2$.

The new physics predicted by the non-maximal mixing model is
$\nu_e\rightarrow\nu_\tau$ oscillations with leading probability (i.e.,
ignoring oscillations involving $\Delta_{sun}$)
\beq
P(\nu_e\rightarrow\nu_\tau)
= 4\,P_{e3} P_{\tau 3} \sin^2 \Delta_{atm} \,,
\label{new}
\eeq
which from Eqs.~(\ref{Ue3}), (\ref{Utau3}), and (\ref{atmetau}) has an
oscillation amplitude
\beq
A^{e\tau}_{atm} = \left[1-\sqrt{1 - A^{e\not e}_{atm}} \, \right]
\left[\sqrt{1 - A^{e\not e}_{atm}} \pm \sqrt{1 - A^{\mu\not\mu}_{atm}}
\, \right] \,,
\label{aetau}
\eeq
where the $\pm$ correlates with the $\mp$ in Eq.~(\ref{Ue3}).
As with $A^{\mu e}_{atm}$, $A^{e\tau}_{atm}$ is completely determined
once $A^{\mu\not\mu}_{atm}$ and $A^{e\not e}_{atm}$ are known, modulo
the two-fold ambiguity in Eq.~(\ref{Umu3}). Contours of
$A^{e\tau}_{atm}$ in the ($A^{\mu\not\mu}_{atm}$,$A^{e\not e}_{atm}$)
plane are shown in Figs.~1(c) and 1(d). We note that the approximate
equalities involving the atmospheric oscillation amplitudes in
Eq.~(\ref{equal}) can also be derived by taking the
$A^{\mu\not\mu}_{atm} \rightarrow 1$, $A^{e\not e}_{atm} \rightarrow 0$
limit in Eqs.~(\ref{amue}) and (\ref{aetau}).

The $\nu_e\rightarrow\nu_\tau$ oscillations could be observed by
long-baseline neutrino experiments with proposed high intensity muon
sources \cite{geer,bww97,bpww}, which can also make precise measurements
of $\nu_\mu\rightarrow\nu_e$, $\nu_\mu\rightarrow\nu_\tau$, and
$\nu_e\rightarrow\nu_\mu$ oscillations. Sensitivity to $A^{e\tau}_{atm}
(\delta m^2_{atm} / {\rm eV}^2)^2 > 2.5\times10^{-9}$ is expected
\cite{geer} for the parameter ranges of interest here. The combination
of all measurements (solar, atmospheric, and long-baseline) would serve
as a consistency check on the model.


{\it Other consequences.} Because the highest mass scale is
$\sqrt{\delta m^2_{atm}} \approx 0.07$~eV, there is little neutrino
contribution to hot dark matter in this model \cite{expansion,hdm}. Such
a mass can be generated for Majorana neutrinos by the see-saw mechanism
with a heavy Majorana scale of about $5\times10^{14}$~GeV (radiative corrections may reduce this
value by about a factor of two \cite{bludman}) if the Dirac
mass is of order the top mass. Also, if $m_1 \ll m_2 < m_3$
in this model, then $m_2/m_3 \sim 10^{-4}$ is approximately equal to
$(m_c/m_t)^2$, as predicted in some see-saw models \cite{bludman}. The
small value of the mass matrix element $M_{\nu_e\nu_e} \approx
10^{-5}$~eV means that the model has an unmeasurably small contribution
to neutrinoless double-beta decay \cite{KK}. The small values of
$U_{e3}$ and $m_3$ mean that there are no observable consequences for
tritium beta-decay \cite{tritium}.  However, the heaviest mass does
provide a relic neutrino target for a mechanism that may generate the
cosmic ray air showers observed above $\gsim 10^{20}$~eV~\cite{relic}.

{\it Discussion.} More precise measurements of the spectrum shape and
day-night effects by SuperK and the Sudbury Neutrino Observatory (SNO)
\cite{SNO} could determine if the solar oscillations are vacuum
long-wavelength. Measurements of a seasonal
variation of the flux of solar $^7$Be and $pep$ neutrinos
\cite{seasonal}, which can be measured in the BOREXINO experiment
\cite{borexino}, would confirm that the solar oscillations are vacuum
long-wavelength. The HELLAZ experiment \cite{HELLAZ} plans to measure
the solar $pp$ neutrino spectrum, which would also be useful in
establishing the vacuum long-wavelength solution. All of
these solar experiments should also be able to pin down the value
$A^{e\not e}_{sun}$ in the vacuum long-wavelength scenario.

Further measurements of atmospheric neutrinos will more precisely
determine the amplitudes $A^{\mu\not\mu}_{atm}$ and
$A^{e\not e}_{atm}$,
completely fixing (modulo a two-fold ambiguity) all oscillation
amplitudes in atmospheric and long-baseline experiments.
The MINOS experiment \cite{MINOS} is expected to test
for $\nu_\mu \rightarrow \nu_e$ and $\nu_\mu \rightarrow \nu_\tau$
oscillations for $\delta m^2_{atm} > 10^{-3}$~eV$^2$. Together the
solar and atmospheric measurements could put strong limits on $U_{e3}$,
which governs the amount of $\nu_e\rightarrow\nu_\tau$ mixing
that could be seen in future long-baseline experiments such as those
discussed for the proposed muon collider at Fermilab \cite{geer}. For
near-maximal solar and atmospheric neutrino mixing the
$\nu_\mu\rightarrow\nu_e$ and $\nu_e\rightarrow\nu_\tau$ oscillation
amplitudes for atmospheric and long-baseline experiments are
approximately equal, and are limited by the deviation of the solar
$\nu_e$ disappearance amplitude from unity.

{\it Acknowledgements.}
We thank John Learned for stimulating discussions on mixing matrices
with maximal mixing and X. Tata for discussions on see-saw mass
matrices.  This work was supported in part by the U.S. Department of
Energy, Division of High Energy Physics, under Grants
No.~DE-FG02-94ER40817, No.~DE-FG05-85ER40226, and No.~DE-FG02-95ER40896,
and in part by the University of Wisconsin Research Committee with funds
granted by the Wisconsin Alumni Research Foundation and the Vanderbilt
University Research Council.

\vfill
\eject

\newpage


\begin{figure}
\centering\leavevmode
\epsfxsize=6.25in\epsffile{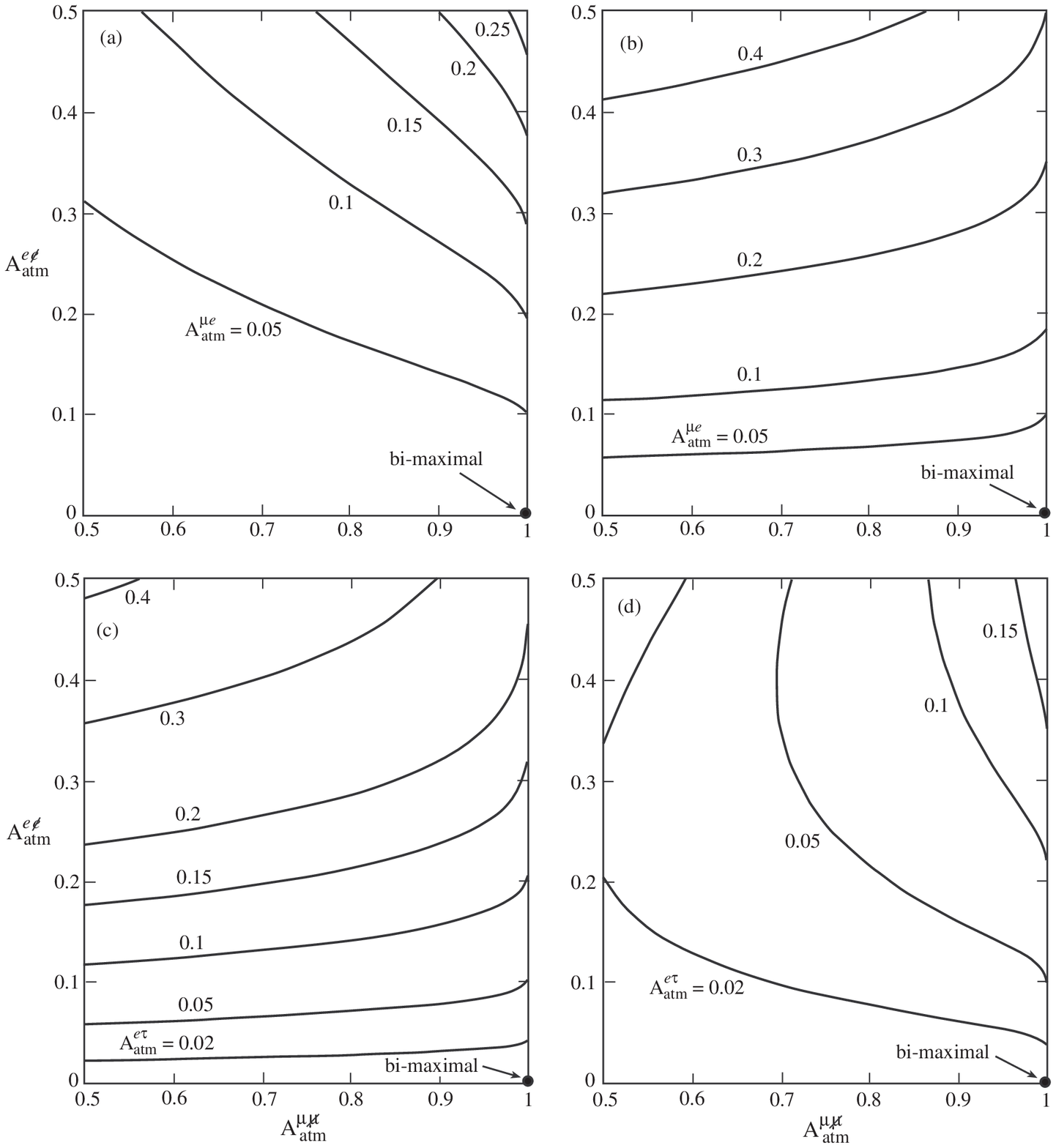}

\bigskip
\caption[]{Correlations of atmospheric oscillation amplitudes: shown are
contours of $A^{\mu e}_{atm}$ in the
($A^{\mu\not\mu}_{atm}$,$A^{e\not e}_{atm}$) plane for the (a)
negative root, and (b) positive root solutions for $P_{\mu3}$ in
Eq.~(\ref{Umu3}), which correspond to $\nu_\mu$ being more closely
associated with $\nu_2$ and $\nu_3$, respectively.
Also shown are contours of $A^{e\tau}_{atm}$ for the (c)
negative root, and (b) positive root solutions. The point with bi-maximal
solar and atmospheric neutrino oscillations is noted.}
\end{figure}

\end{document}